\documentclass[showpacs, preprintnumbers, twocolumn]{revtex4}
\usepackage[dvips]{graphicx}
\usepackage{dcolumn}
\usepackage{bm}

\begin{document}

\title{Experimental observation of the Bogoliubov transformation for a
Bose-Einstein condensed gas}

\begin{abstract}
Phonons with wavevector $q/\hbar $ were optically imprinted into a
Bose-Einstein condensate. Their momentum distribution was analyzed using
Bragg spectroscopy with a high momentum transfer. The wavefunction of the
phonons was shown to be a superposition of $+q$ and $-q$ free particle
momentum states, in agreement with the Bogoliubov quasiparticle picture.

%
\end{abstract}

\author{J.M. Vogels}
\author{K. Xu}
\author{C. Raman}
\altaffiliation{Current Address: School of Physics
Georgia Institute of Technology, Atlanta, GA 30332}
\author{J. R. Abo-Shaeer}
\author{W. Ketterle}
\homepage[Group website: ]{http://cua.mit.edu/ketterle_group/}
\affiliation{Department of Physics, MIT-Harvard Center for Ultracold
Atoms, and Research Laboratory of Electronics,
Massachusetts Institute of Technology, Cambridge, MA 02139}
\maketitle%
%
\catcode`\ä = \active \catcode`\ö = \active \catcode`\ü = \active
\catcode`\Ä = \active \catcode`\Ö = \active \catcode`\Ü = \active
\catcode`\ß = \active \catcode`\é = \active \catcode`\è = \active
\catcode`\ë = \active \catcode`\ô = \active \catcode`\ê = \active
\catcode`\ø = \active \catcode`\ò = \active \catcode`\í = \active
\defä{\"a} \defö{\"o} \defü{\"u} \defÄ{\"A} \defÖ{\"O} \defÜ{\"U} \defß{\ss} \defé{\'{e}}
\defè{\`{e}} \defë{\"{e}} \defô{\^{o}} \defê{\^{e}} \defø{\o} \defò{\`{o}} \defí{\'{i}}%
%
The pioneering work of Bogoliubov in 1947 constitutes the first microscopic
theory that attributes superfluidity to Bose-Einstein condensation \cite
{bogo47}. As described by Einstein noninteracting bosons condense \cite
{eins25quan}, but they are not superfluid. However, Bogoliubov showed that
with weak interactions the condensate will exhibit superfluidity. Repulsive
interactions change the elementary excitations at long wavelengths from free
particles into phonons which, according to the Landau criterion, leads to
superfluidity \cite{land41}. The main step in the non-perturbative treatment
is the Bogoliubov transformation 
\begin{eqnarray}
b_{+q}^{\dagger } &:&=u_{q}a_{+q}^{\dagger }+v_{q}a_{-q}  \label{bogo} \\
b_{+q} &:&=u_{q}a_{+q}+v_{q}a_{-q}^{\dagger }  \nonumber
\end{eqnarray}
which expresses the creation and annihilation operators $b_{q}^{\dagger }$
and $b_{q}$ for Bogoliubov quasiparticles in terms of creation and
annihilation operators $a_{q}^{\dagger }$ and $a_{q}$ for free particles in
momentum states $q$. This transformation also plays a crucial role in
general relativity, where it connects the particle operators in different
reference frames~\cite{bire82}. In this paper we experimentally verify the
Bogoliubov transformation by generating quasiparticles with momentum $+q$
and observing that they are a superposition of $+q$ and $-q$ momentum states
of free particles.

Following a recent suggestion of Brunello et al. \cite{brun00}, we perform
an experiment that combines the two regimes of Bragg spectroscopy, where the
momentum imparted to the atoms is either smaller \cite{stam99phon} or larger 
\cite{sten99brag} than the sound velocity $c$ times the mass $m$ of the
atoms. An optical lattice moving through the condensate at the sound
velocity imprinted phonons with wavelengths equal to the spatial period of
the lattice \cite{stam99phon}. This was accomplished by intersecting two
laser beams at a small angle and choosing their frequency difference to be
equal to the phonon frequency. The momentum analysis of the phonon
wavefunction was performed by a second Bragg pulse consisting of
counterpropagating laser beams that transferred a large momentum $Q$ (two
photon recoils) to atoms with initial momentum $p$. In the limit $Q\gg p$, $%
mc$, the resonance frequency $\nu $ is equal to the kinetic energy
transferred: $h\nu =Q^{2}/2m+Qp/m$, where the second term is simply the
Doppler shift\cite{sten99brag,zamb00,blak01bragg}. The resulting frequency
spectrum shows three peaks corresponding to the three momentum components ($0
$, $+q$ and $-q$) of a condensate with phonons (see Fig. \ref{fig:sidebands}%
) \cite{brun00}.

The Bogoliubov spectrum for the energy of elementary excitations is \cite
{huan87} 
\begin{equation}
\varepsilon (q)=\sqrt{q^{2}c^{2}+\left( \frac{q^{2}}{2m}\right) ^{2}}
\label{dispersion}
\end{equation}
where $c=\sqrt{4\pi \hbar ^{2}an}/m$, $a$ is the scattering length, and $n$
is the density. The amplitudes $u_{q}$ and $v_{q}$ in Eq. \ref{bogo} are
given by 
\begin{equation}
u_{q},v_{q}=\pm \frac{\varepsilon (q)\pm \frac{q^{2}}{2m}}{2\sqrt{
\varepsilon (q)\frac{q^{2}}{2m}}}\text{.}  \label{uv}
\end{equation}
The non-trivial aspect of the Bogoliubov transformation manifests itself
only at low momenta ($q\ll mc)$, where both amplitudes $u_{q}$ and $v_{q}$
are significant. In this regime, the excitations are characterized by $u\sim
-v\sim \sqrt{mc/2q}\gg 1$ and $\varepsilon (q)\sim qc$. Such excitations are
phonons, each involving many particles moving in both directions. At high
momenta ($q\gg mc$), $u_{q}\sim 1$, $v_{q}\sim 0$ and $\varepsilon
(q)=q^{2}/2m+mc^{2}$. These excitations are free particles with an energy
shift equal to the chemical potential $\mu =mc^{2}$.

The phonon creation operator $b_{+q}^{\dagger }$ (Eq. \ref{bogo}) is a
superposition of the creation operator $a_{+q}^{\dagger }$ for particles
moving in the $+q$ direction and the $\emph{annihilation}$ operator $a_{-q}$
for particles moving in the $-q$ direction, yet the creation of a phonon in
the condensate implies an $\emph{increase}$ of particles moving in both the $%
+q$ and $-q$ direction. Indeed, simple operator algebra shows that a
condensate with $l$ excitations $b_{+q}^{\dagger l}/\sqrt{l!}\left| \Psi
_{0}\right\rangle $ contains $lu_{q}^{2}+v_{q}^{2}$ free particles moving
with momentum $+q$ and $(l+1)v_{q}^{2}$ free particles with momentum $-q$.
In its ground state the condensate contains $v_{q}^{2}$ pairs of atoms with
momenta $+q$ and $-q$. These pairs constitute the quantum depletion in the
condensate wavefunction \cite{huan87}
\begin{equation}
\left| \Psi _{0}\right\rangle =\prod_{q\neq 0}\frac{1}{u_{q}}%
\sum_{j=0}^{\infty }\left( -\frac{v_{q}}{u_{q}}\right) ^{j}\left|
n_{-q}=j,n_{+q}=j\right\rangle \text{.}
\end{equation}
where the remaining atoms are in the $q=0$ momentum state. When $%
a_{+q}^{\dagger }$ and $a_{-q}$ act on $\Psi _{0}$, terms with large
occupation numbers $j$ are enhanced: $a_{+q}^{\dagger }\left|
n_{-q}=j,n_{+q}=j\right\rangle =\sqrt{j+1}\left|
n_{-q}=j,n_{+q}=j+1\right\rangle $, $a_{-q}\left|
n_{-q}=j,n_{+q}=j\right\rangle =\sqrt{j}\left|
n_{-q}=j-1,n_{+q}=j\right\rangle $. In addition, $-q$ and $+q$ atoms appear
only in pairs. Together, these two effects cause both $a_{+q}^{\dagger }$
and $a_{-q}$ to increase the number of atoms in both the $-q$ and $+q$
states.

The experiments were performed with condensates of $3\times 10^{7}$ sodium
atoms in a magnetic trap with radial and axial trapping frequencies of 37 Hz
and 7 Hz, respectively~\cite{onof00}. The condensate had a peak density of $%
1.0\times 10^{14}$ cm$^{-3}$ corresponding to a chemical potential of $\mu
=h\cdot $1.5 kHz, a sound velocity of $c$ = 5 mm/s, and a Thomas-Fermi
radial radius of 32 $\mu $m. The Bragg beams for the optical lattices were
generated from a common source of laser light 1.7 GHz red-detuned from the $%
3S_{1/2}$ $\left| F=1\right\rangle $ to $3P_{3/2}$ $\left| F^{\prime
}=0,1,2\right\rangle $ transitions. The lattices were moved radially through
the cloud. This was done to avoid the high collisional density along the
axial direction, where outcoupled atoms would have undergone elastic
collisions with the condensate \cite{chik00}. The Bragg beams for imprinting
the phonons were at an angle of 32 mrad with respect to one another,
resulting in a lattice spacing of $\sim $ 9 $\mu $m and a recoil momentum of 
$q$ = $m\cdot$1.9 mm/s. The beams propagated at an angle of 0.4 rad with
respect to the longitudinal axis of the condensate and were linearly
polarized perpendicular to it. The frequency difference (``excitation
frequency'') between the two beams was chosen to be 400 Hz, corresponding to
the frequency of the phonons. The excitation pulse had to be long enough to
ensure sufficient frequency resolution in order to selectively excite $+q$
phonons, and no $-q$ phonons (see below). However, the excitation pulse also
had to be shorter than the transit time of phonons through the condensate,
since phonons accelerate when they move through regions of varying density.
We chose small-angle Bragg beams 3 ms in duration with intensity 0.05 mW/cm$%
^{2}$, corresponding to a two-photon Rabi frequency of 50 Hz.

\begin{figure}[tbp]
\includegraphics{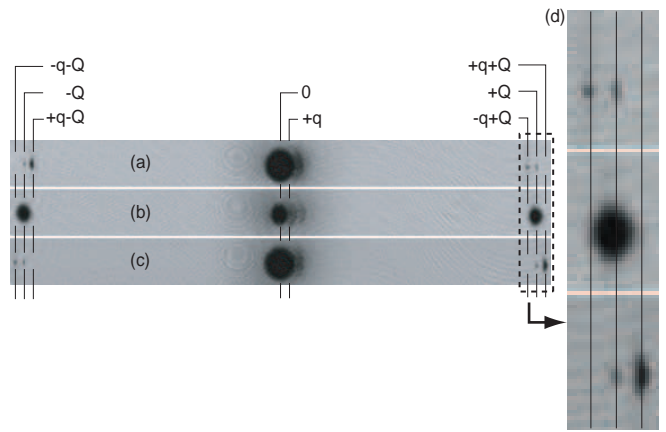}
\caption{Momentum distribution of a condensate with phonons. After
imprinting $+q$ phonons into the condensate, momentum analysis via Bragg
spectroscopy transfers a momentum $\pm Q$ (two photon recoil) to the atoms.
Absorption images after 40 ms time of flight in (a), (b), and (c) show the
condensate in the center and outcoupled atoms to the right and left for
probe frequencies of 94, 100, and 107~kHz, respectively. The small clouds
centered at $+q$ are phonons that were converted to free particles. The size
of the images is 25$\times $2.2 mm. (d) The outlined region in (a) - (c) is
magnified, and clearly shows outcoupled atoms with momenta $Q\pm q$,
implying that phonons with wavevector $q/\hbar $ have both $+q$ and $-q$
free particle momentum components.}
\label{fig:abs-images}
\end{figure}

\begin{figure}[tbp]
\includegraphics[height=2in]{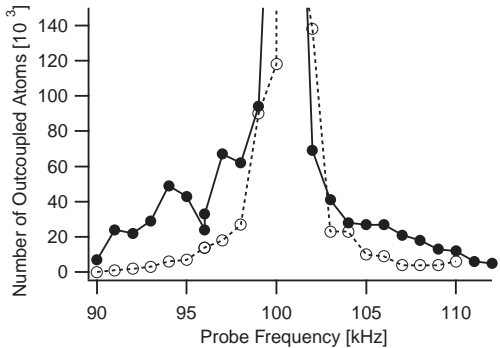}
\caption{Bragg spectrum of a condensate with ($\bullet$) and without ($\circ$%
) phonon excitation. The number of outcoupled atoms vs. the probe frequency
is shown. The excitation of phonons creates sidebands on both sides of the
central condensate peak. Around 100 kHz, the absorption images were
saturated due to high optical density.}
\label{fig:sidebands}
\end{figure}

The momentum analysis of the phonons was performed with two
counterpropagating beams that imparted a recoil momentum of $Q$ = $m\cdot $%
59 mm/s onto the outcoupled atoms. The beams were polarized parallel to the
longitudinal axis of the condensate to suppress superradiant emission\cite
{inou99super}. A frequency difference (``probe frequency'') of 100 kHz
between the two beams corresponded to the kinetic energy needed for atoms
initially at rest to reach this recoil momentum. To gain recoil momentum $+Q$%
, atoms with initial momentum $+q$ were resonant at 107 kHz (Fig. \ref
{fig:abs-images}a). Atoms with momentum $-q$ were resonant at 94 kHz (Fig. 
\ref{fig:abs-images}b). In our experiment, a retroreflected beam containing
both optical frequencies resulted in two optical lattices moving at the same
speed in opposite directions. This led to simultaneous outcoupling of $+q$
and $-q$ atoms in opposite directions. These large-angle Bragg beams were
pulsed on for 0.5 ms. The probe pulse had to be long enough to selectively
excite atoms with $\pm q$ momentum, but also shorter than $h/\mu $ for the
mean field energy to be negligible during the readout. Each frequency
component had an intensity of 0.7 mW/cm$^{2}$, corresponding to a two-photon
Rabi frequency of 700 Hz. Subsequently, the trap was turned off and a
resonant absorption image was taken after 40 ms of ballistic expansion.

Fig.~1 shows typical absorption images for various probe frequencies. The
quasiparticle nature of the phonons was directly evident (see outlined
region) in the time-of-flight distribution through the presence of the peaks
at momenta $\pm q+Q$. These peaks had well defined momentum because the
outcoupled atoms left the condensate quickly (during this time the atoms'
velocity changed by less than the speed of sound \cite{hagl99}). This
``photograph'' of the Bogoliubov transformation is the central result of
this paper.

We now discuss the different momentum components distinguishable in Fig.~1:

(1) The original condensate is in the center.

(2) The condensate is asymmetrically extended towards positive momenta. This
is due to imprinted phonons of momentum $q$ that were converted to free
particles during the ballistic expansion. Previously this signature was used
to determine the structure factor of a condensate~\cite{stam99phon}. There
is no momentum component in the opposite direction, because we did not
create $-q$ phonons (see below). The smearing of the observed momentum
distribution may be caused by acceleration due to the inhomogeneous density
distribution of the condensate when the trap was switched off.

(3) The two components with momentum $+Q$ and $-Q$ are atoms outcoupled from
the condensate at rest. The symmetry of the $\pm Q$ peaks, as well as the
position of the main condensate, served as an indicator as to whether the
condensate was undergoing dipole oscillation during the experiment. Images
that showed such ``sloshing'' (caused by technical noise) were excluded from
further analysis.

(4) Atoms at $+q+Q$ (Fig.~1c) and $+q-Q$ (Fig.~1a) are coupled out from the $%
+q$ component of the phonons.

(5) Atoms at $-q-Q$ (Fig.~1c) and $-q+Q$ (Fig.~1a) are coupled out from the $%
-q$ component of the phonons.

Quantitative information was obtained by scanning the probe frequency and
measuring the number of outcoupled atoms in the three peaks around $-Q$ (See
Fig.~1). Without phonons only the condensate peak is observed. The
excitation of phonons at wavevector $+q$ creates momentum sidebands at
momentum $\pm q$. The $-q$ peak is expected to be smaller by a factor $%
v_q^{2}/u_q^{2}$. This effect is evident in Fig.~2, but with a poor
signal-to-noise ratio.

The excitation of the phonons was characterized by scanning the frequency
difference of the small angle Bragg beams (Fig.~3a) and measuring the number
of outcoupled atoms at $+q-Q$. The probe frequency was kept at 94 kHz
because the momenta of the excited phonons remained fixed. The observation
of two distinct peaks (corresponding to $+q$ at 400 Hz and $-q$ phonons at
-400 Hz, respectively) confirms that there was sufficient resolution to
excite only $+q$ phonons and suppress the off-resonant excitation of $-q$
phonons. The higher peak represents the $u$-component of $+q$ phonons
excited at positive excitation frequencies. When the excitation frequency
became negative, $-q$ phonons were excited. This second peak represents the $%
v$-component of the phonons.

According to Eq.~\ref{uv} the ratio of the two peaks, $v_{q}^{2}/u_{q}^{2}$,
should be smaller at lower density. This is confirmed in Fig.~3b, which
shows the excitation spectrum at low density. The density was lowered by a
factor of 2 by weakening the axial trap frequency to 4 Hz and reducing the
number of atoms in the condensate by a factor of 3. The scatter in the data
could be due to residual sloshing, shot-to-shot fluctuations in the size of
the condensate, or non-linear effects (see below).

\begin{figure}[tbp]
\includegraphics[height=4in]{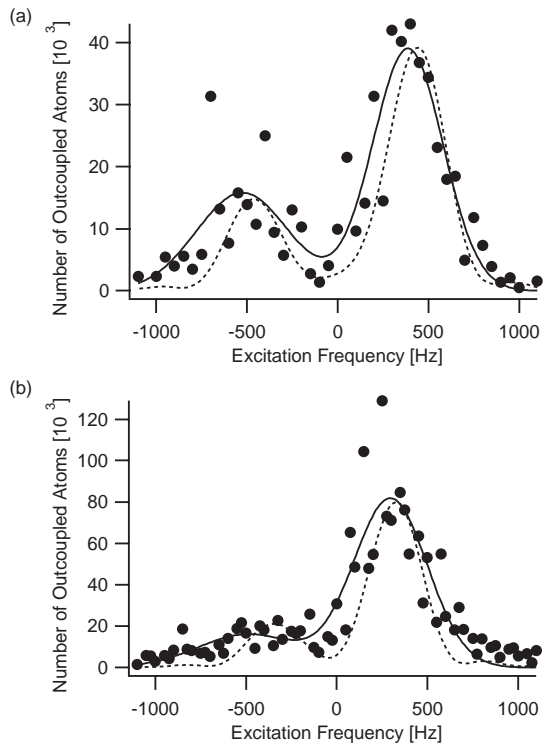}
\caption{Phonon excitation spectrum. Atoms with initial momentum $+q$ were
detected by setting the probe frequency to $94$ kHz and measuring the number
of atoms with momentum $-Q+q$. The two peaks reflect that phonons with $%
+q/\hbar $ and $-q/\hbar $ wavevectors have free particle components with
momentum $+q$. Spectra were taken at: (a) high density (1.0$\times $10$^{14}$
cm$^{-3}$) and (b) low density (0.5$\times $10$^{14}$ cm$^{-3}$). The solid
line, a fit to the sum of two Gaussians, is intended to guide the eye. The
dashed line is the theoretical prediction.}
\label{fig:exc-spectrum}
\end{figure}

The dashed line is a the theoretical prediction with no free parameters
(except the vertical scale). It was obtained by integrating Eq.~\ref{uv}
over the inhomogeneous density distribution of the condensate and by
accounting for the finite length of the square excitation pulse, which
broadens the line and causes the extra sidelobes around 900 Hz. The theory
assumes the validity of a perturbative approach, i.e. that the excitation
pulse is weak.

Ideally, both Bragg pulses should affect only a few percent of the atoms, as
in previous experiments \cite{stam99phon,sten99brag}. However, because our
signal was the product of the two outcoupling efficiencies, we needed to
work at much higher outcoupled fractions. We estimate that the excitation
pulse transferred 10\ of the condensate atoms into the phonon state, and
that the probe pulse outcoupled 40\% of these atoms on resonance.

Both Bragg processes (stimulated Rayleigh scattering) should depend only on
the product of the two intensities. Therefore, changing the sign of the
excitation frequency should not affect the number of phonons generated
(assuming a stationary condensate). However, when the sign of the excitation
frequency was changed in our experiment, the observed number of phonons
differed. This is most likely due to superradiance \cite{inou99super}, which
is sensitive to the individual intensities. Therefore, the asymmetry in
phonon number was eliminated by ensuring equal intensities in the two beams.
Additionally, there was a substantial loss ($\sim $50\%) of condensate atoms
due to superradiant Rayleigh scattering. Both superradiant effects could be
further suppressed using light further detuned from resonance. However, in
our case this would have required an additional laser. Given these
experimental limitations, the agreement between experiment and theory in
Fig. 3 is satisfactory.

In this work we have used large-angle Bragg pulses to analyze the momentum
structure of the phonon wavefunction. In principle, this could have also
been achieved by removing the mean-field interaction within a time $%
h/\varepsilon (q)$ and then probing the velocity distribution of the
particles. This is not possible with ordinary ballistic expansion because
the reduction of the interactions is too slow, taking place on a time scale
of the trapping period. However, the use of a Feshbach resonance \cite
{inou98} would provide an effective method for suddenly reducing the
mean-field interaction.

In conclusion, we have experimentally analyzed the phonon wavefunction in a
Bose-Einstein condensate. Following recent theoretical work~\cite{brun00},
the two-component character of Bogoliubov quasi-particles was observed in
the frequency domain (Fig.~2). In addition, the momentum components of the
phonon wavefunction were discriminated by their final momenta after the
probe pulse (Fig.~1). By combining momentum and frequency selectivity, we
were able to directly ``photograph'' the Bogoliubov transformation
(Fig.~1d), demonstrating the power of Bragg spectroscopy to analyze
non-trivial wavefunctions. This method may be also applicable to studying
the many-body and vortex states \cite{zamb00} of dilute atomic Bose-Einstein
condensates.

This work was funded by ONR, NSF, ARO, NASA, and the David and Lucile
Packard Foundation. We are grateful to A. Brunello and S. Stringari for
insightful discussions.

\end{document}